\documentclass{article}
\usepackage{psfig}
\usepackage{graphicx}

\def\lsi{\raise0.3ex\hbox{$<$\kern-0.75em\raise-1.1ex\hbox{$\sim$}}}
\def\gsi{\raise0.3ex\hbox{$>$\kern-0.75em\raise-1.1ex\hbox{$\sim$}}}

\newcommand{\gz}{\frac{GM}{\Omega^2 r^3}}
\newcommand{\gzt}{{GM}/{\Omega^2 r^3}}
\newcommand{\rco}{\left(\frac{r_{\rm co}}{r}\right)^3}
\newcommand{\lfp}{\left(}
\newcommand{\rgp}{\right)}
\newcommand{\beq}{\begin{equation}}
\newcommand{\eeq}{\end{equation}}

\begin{document}

\title{Distribution of magnetically confined circumstellar matter in oblique rotators}

\author{O.~Preuss, M.~Sch\"ussler, S.K. Solanki \thanks{mail to: schuessler@linmpi.mpg.de}\\
 Max-Planck-Institut f\"ur Aeronomie \\ 37191 Katlenburg-Lindau, Germany \\[0.5cm]
V.~Holzwarth \\School of Physics and Astronomy\\
	University of St. Andrews \\ St. Andrews Fife KY16 9SS, Scotland, UK}

\maketitle

\begin{abstract}  
  We consider the mechanical equilibrium and stability of matter trapped
  in the magnetosphere of a rapidly rotating star. Assuming a dipolar 
  magnetic field and arbitrary inclination of the magnetic axis with 
  respect to the axis of rotation we find stable equilibrium positions 
  a) in a (warped) disk roughly aligned with the magnetic equatorial 
  plane and b) at two locations above and below the disk, whose distance 
  from the star increases with decreasing inclination angle between dipole 
  and rotation axis. The distribution of matter is not strongly affected 
  by allowing for a spatial offset of the magnetic dipole. These results 
  provide a possible explanation for some observations of corotating localized 
  mass concentrations in hot magnetic stars.

\end{abstract}

  \section{Introduction}
    Spectroscopic observations indicate the presence of corotating
    circumstellar matter around chemically peculiar stars
    (helium-strong, helium-weak, and Ap/Bp stars) with strong
    magnetic fields (e.g., Shore \& Adelman 1981, Walborn 1982, Shore
    et al. 1990, Groote \& Hunger 1997, Smith et al. 1998, Smith
    2001). Such accumulation of mass often coexists with magnetically
    confined outflows (Barker et al. 1982, Brown et al. 1985, Shore et
    al. 1987, Smith \& Groote 2001). Shore and Brown (1990, see also
    Shore 1987) suggested that the corotating plasma is trapped in a
    magnetically closed region near the magnetic equator while the
    stellar wind forms jet-like outflows along open field lines around
    the magnetic poles. This is similar to the earlier concept of
    Mestel (1968, see also Mestel \& Spruit 1987) of `dead zones' and
    `wind zones' and also reminiscent to the structure of the solar
    corona during activity minimum (e.g., Pneuman \& Kopp 1971).
    Recently, Ud-Doula \& Owocki (2002) have carried out two-dimensional
    MHD simulations illustrating the development of open and closed 
    field topologies around a non-rotating hot star with a line-driven 
    wind.

    A mechanism for the accumulation of circumstellar matter in
    equatorial disks around Be stars has been proposed by Bjorkman \&
    Cassinelli (1993, see also Cassinelli et al. 2002, and references
    therein). They suggest that the strong magnetic field forces the
    radiatively driven outflowing matter to follow the dipolar field
    lines and collide near the magnetic equator.  Since the colliding
    flows are supersonic, shock heating leads to UV and X-ray
    emission. After the material has cooled sufficiently, it accumulates
    near the magnetic equator and forms a disk. Babel and Montmerle
    (1997) have proposed a similar mechanism for corotating
    circumstellar matter and X-ray emission from Ap/Bp stars. 
    For the matter to become magnetically trapped in a closed part of
    the magnetosphere, the kinetic energy density of the wind has to be
    smaller than the magnetic energy density. Depending on the strength
    of the magnetic field, wind speed and mass loss rate, estimates
    yield an extension of the corotating magnetosphere of up to 10
    stellar radii.

    Most of the previous theoretical considerations of the equilibrium
    and dynamics of circumstellar matter around magnetic stars
    considered {\em aligned} rotators, for which the magnetic axis and
    the rotation axis coincide. In this case, the trapped matter
    accumulates near the equatorial plane. The
    distribution of the magnetospheric matter in an {\em oblique} rotator with
    arbitrary relative inclination of the two axes is far less clear. 
    Here we consider this general case by studying the force
    equilibrium and the stability of circumstellar matter in the closed
    magnetosphere of an oblique rotator. Assuming that the
    magnetic energy density is much larger than the kinetic energy
    density of rotation, which, in turn, is much larger than the thermal
    energy density of the plasma, we can assume the magnetic field to be
    fixed and ignore the thermal pressure. The equilibrium problem is then
    reduced to finding the locations where the (vector) sum of the
    gravitational and the centrifugal force is perpendicular to a given
    magnetic field line. A similar approach has been taken by Nakajima
    (1985), who considered the minima of the effective
    potential energy along a given magnetic field
    line. Nakajima apparently determined the absolute minima of the
    potential and thus could only find one equilibrium locus per field
    line. In our work, however, we show that there are more locations
    where the trapped plasma can accumulate. We thus are able to predict
    the complete spatial distribution of the magnetospheric plasma in a
    rapidly rotating magnetic star.

    The paper is organized as follows. Sec.~\ref{sec_equil} describes
    the method to determine the force equilibrium and its
    stability. Sec.~\ref{sec_results} gives results for the special
    cases of aligned and perpendicular rotators as well as for other
    relative inclinations of the rotation and magnetic axes. We also
    consider offset dipoles. We discuss the results in
    Sec.~\ref{sec_disc} and present our conclusions in Sec.~\ref{sec_conc}.
      
  \section{Force equilibrium and stability}
  \label{sec_equil}

    We consider the closed part of the magnetosphere of a rapidly
    rotating star, i.e., the region inside the Alfv\'en radius defined
    by the equality of stellar wind speed and Alfv\'en velocity. We
    assume the scaling
    \begin{equation}
    v^2_{\rm s} \ll v^2_{\Omega} \ll v^2_{\rm A}\,,
    \label{eq_rel}
    \end{equation}
    where $v_{\rm s}$, $v_{\Omega}$, and $v_{\rm A}$ are the local sound
    speed, rotational speed (in an inertial frame), and Alfv\'en speed, 
    respectively. This scaling corresponds to a situation where the 
    magnetic energy density is much larger than the kinetic energy density 
    of the rotational motion, which, in turn, is much larger than the thermal 
    energy density of the magnetospheric matter. Under these conditions, which
    are relevant for the observed stellar magnetospheres out to about 10
    stellar radii (Nakajima 1985; Shore 1987; Babel \& Montmerle
    1997), the magnetic field is largely unaffected by the presence of
    circumstellar matter. In the limit of very large electrical 
    conductivity, the gas is attached to the magnetic field lines and
    the distribution of the circumstellar matter along a given field
    line is determined by the gravitational and the centrifugal forces
    (in a corotating frame of reference). Since under the scaling given by
    Eq.~(\ref{eq_rel}) the field lines are
    practically rigid (a minimal distortion of the field can balance any
    force perpendicular to the field lines), the force equilibrium for a
    mass element $\delta m$ is given by the balance of the components of the
    gravitational and centrifugal forces tangential to the local field
    line,

    \beq\label{equi}
     ({\bf F}_{\rm C}+{\bf F}_{\rm G})\cdot {\bf B}=0 \; ,
    \eeq
    where ${\bf B}$ is the magnetic field vector,
    \beq\label{fz}
      {\bf F}_{\rm C} = \delta m\,\, \Omega^2 r 
                  \;\left[{\bf e}_r-{\bf e}_\Omega\, 
                  ({\bf e}_r \cdot {\bf e}_\Omega)\right] 
    \eeq
    is the centrifugal force and 
    \beq\label{fg}            
      {\bf F}_{\rm G} = - \,G \frac{M \delta m}{r^2}\;{\bf e}_r       
    \eeq
    is the gravitational force.
    Here ${\bf e}_r$ denotes the radial unit vector, pointing from the
    center of the star to the mass element located at a radial distance
    $r$. ${\bf e}_\Omega$ is the unit vector along the axis of
    rotation, $G$ denotes the gravitational constant and $M$ the mass
    of the star. We consider a magnetic dipole field with a
    dipole moment, ${\bf m}$, parallel to the unit vector ${\bf e}_{\rm m}$:
    \beq\label{bdp} {\bf B} = \frac{\mu_0}{4 \pi
      r^3}\; \left[3({\bf e}_r \cdot {\bf m}) {\bf e}_r-{\bf
      m}\right] \; .
    \eeq
    Inserting Eqs.~(\ref{fz}) - (\ref{bdp}) in Eq.~(\ref{equi}) yields
    \beq 
      \left[{\bf e}_r\lfp1-\gz\rgp-{\bf e}_\Omega\,({\bf e}_r \cdot
      {\bf e}_\Omega)\right] \cdot\biggr[3({\bf e}_r \cdot{\bf e}_{\rm m})
      {\bf e}_r-{\bf e}_{\rm m} \biggl]=0 \; ,
    \eeq
    which can be written as
    \beq\label{equi1} 
      A\,({\bf e}_r \cdot{\bf e}_{\rm m})+ ({\bf e}_r
      \cdot {\bf e}_\Omega)({\bf e}_\Omega \cdot{\bf e}_{\rm m}) = 0 \;  
     \eeq
     with
     \beq\label{equi8}
       A = 2\left(1-\gz\right)-3({\bf e}_r \cdot {\bf e}_\Omega)^2\; .
     \eeq
     The quantity $\gzt$ represents the ratio of the gravitational to
     the centrifugal force. Kepler rotation would give a value of unity for
     this quantity. $\left(GM/\Omega^2\right)^{1/3}\equiv r_{{\rm co}}$ is 
     thus called the {\it corotation radius}.
     
    Eqs.~(\ref{equi1}) - (\ref{equi8}) show that the distribution of the 
    equilibrium positions is scale-invariant for a fixed value of the angle
    $\psi$ between the magnetic and the rotational axes $({\bf e}_\Omega
    \cdot{\bf e}_{\rm m}=\cos\psi)$. Therefore, writing $r$ in units of
    $r_{\rm co}$ allows us to obtain the result for any values of the stellar
    parameters $\Omega$ and $M$ after proper rescaling with $r_{\rm co}$.
    
    To analyse the stability of a mass element located at an equilibrium
    position ${\bf r}_0$ we consider the projection of the total force
    $({\bf F}={\bf F}_{\rm C}+{\bf F}_{\rm G})$ on the tangent vector
    ${\bf l}$ of the field line for a small tangential displacement,
    $\eta$, of a mass element from its equilibrium position along the
    corresponding magnetic field line. Denoting the equilibrium quantities 
    by a subscript zero and neglecting second- and higher-order contributions
    in $\eta$, we have

    \begin{eqnarray}\label{st1}
      {\bf F}\cdot{\bf l}&=&({\bf F}_0+\delta{\bf F})\cdot({\bf l}_0+
             \delta{\bf l})\\\nonumber
      &=&[{\bf F}_0+\eta\,({\bf l}_0\cdot{\bf \nabla}){\bf F}({\bf r_0})]
             \cdot({\bf l}_0+{\bf n}_0\,\eta\,\kappa^{-1}_0)\;,
    \end{eqnarray} 
    with
    \beq
      {\bf F}({\bf r}_0+\eta\,{\bf l}_0)={\bf F}_0+\eta\,
             ({\bf l}_0\cdot{\bf \nabla}){\bf F}({\bf r_0})
     \eeq
    and
    \begin{eqnarray}
      {\bf l}({\bf r}_0+\eta\,{\bf l}_0)&=&{\bf l}_0+\eta\,
              ({\bf l}_0\cdot\nabla){\bf l}({\bf r}_0)\\
      \nonumber&=&{\bf l}_0+\eta\,\kappa^{-1}_0\,
                  {\bf n}_0 \; .
    \end{eqnarray}
    $\kappa_0$ is the local curvature radius and ${\bf n}_0$ is the normal 
    vector to the field line through ${\bf r}_0$. From Eq.~(\ref{st1}) we obtain
    \beq
      {\bf F}\cdot{\bf l}=\left\{\left[({\bf l}_0\cdot{\bf \nabla}){\bf F}
             \right]\cdot{\bf l}_0+
      {\bf F}_0\cdot{\bf n}_0\,\kappa^{-1}_0\right\}\eta 
    \eeq
    to first order in $\eta$, where we have used the equilibrium condition
    ${\bf F}_0 \cdot {\bf l}_0 = 0$.
    The neccessary and sufficient condition for a stable equilibrium is
    that the projection of ${\bf F}$ on
    ${\bf l}$ is directed towards the equilibrium position, viz.
    \beq\label{stbeq}
      \left[({\bf l}_0\cdot{\bf \nabla}){\bf F}\right]
      \cdot{\bf l}_0+
      {\bf F}_0\cdot{\bf n}_0\,\kappa^{-1}_0 < 0\;.
    \eeq

\section{Results}
\label{sec_results}
      \begin{figure*}[t]
        \centerline{\psfig{figure=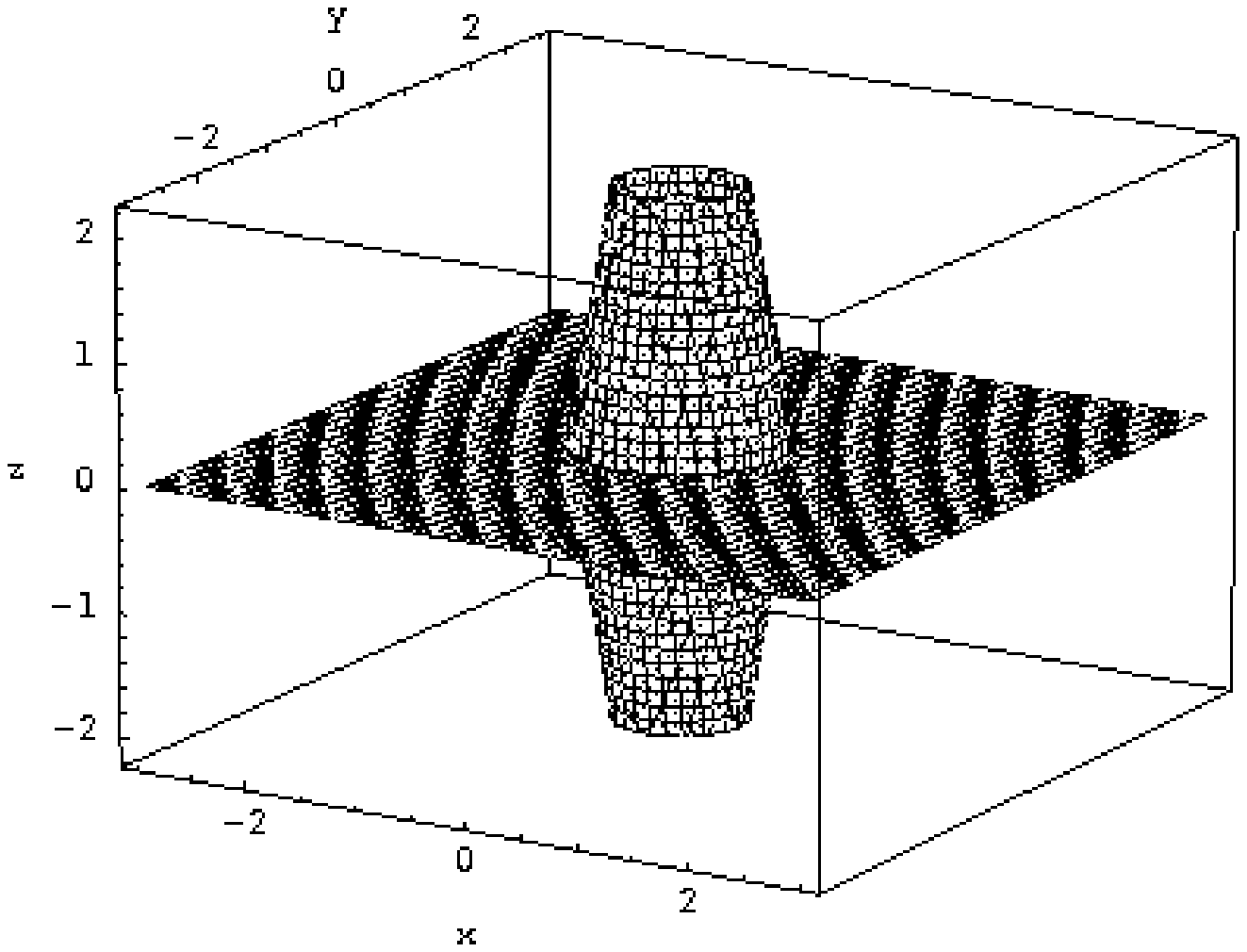,height=2.5in,width=3.2in}
	            \psfig{figure=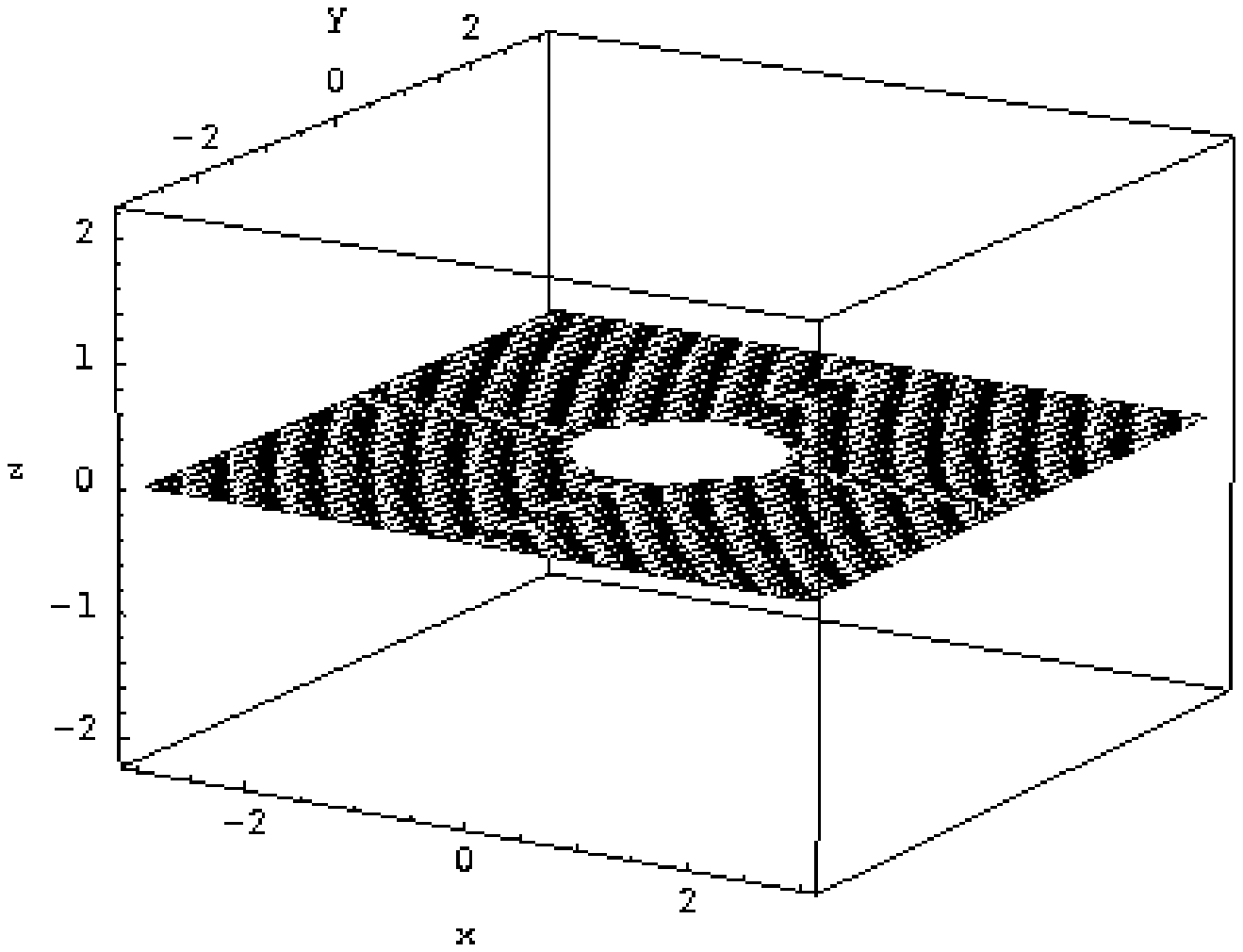,height=2.5in,width=3.2in}}
        \caption{Hatched surfaces show the equilibrium positions for the  
	 case $\psi=0$ (aligned rotator)
	 Both the magnetic dipole axis and the rotation axis are in $z$-direction. The 
	 left panel shows all equilibrium positions while the right panel shows 
	 only the stable ones. All lengths are given in units of the corotation radius.}
	\label{fig_aligned}         
      \end{figure*} 
    
    For the following illustrative examples we use parameter values 
    corresponding to the A0p star IQ Aur (HD 34452, see Babel 
    and Montmerle 1997), namely $M =4.8\,M_{\odot}$, $R_{\star}=5\, 
    R_{\odot}$ and $P_{\rm rot}=2.5\, {\rm d} \rightarrow\Omega= 2.9\cdot 
    10^{-5}\,{\rm s}^{-1}$ and $r_{\rm co}=13.4\,R_{\odot}=2.6\,R_{\star}$. 
    Note that the results for cases with other values of $M$ and $\Omega$ 
    can simply be found by rescaling with the corresponding value of $r_{\rm co}$,
    so that the results obtained here are, in general, valid for other stellar 
    parameters as well. 
    We first consider the special cases of aligned and perpendicular
    rotators and then show results for the oblique case.
  
   \subsection{Aligned rotator: $\psi=0$}
   
      From Eq.~(\ref{equi1}) we find for the case of parallel rotation and
      magnetic axes $({\bf e}_{\rm m} \cdot {\bf e}_\Omega = 1)$ the
      equilibrium condition
      \beq
        \left\{2\left[1-\rco\right]-3({\bf e}_r \cdot {\bf
        e}_\Omega)^2 + 1\right\} ({\bf e}_r \cdot {\bf e}_\Omega) =0\;,
      \eeq 
      with the rotation axis directed along the $z$-axis.
      This equation has two solutions:
      \begin{itemize}
        \item[1)] $\cos\theta\equiv{\bf e}_r \cdot {\bf e}_\Omega=0$:
                 this solution corresponds to the (coinciding magnetic
                 and rotational) equatorial plane of the star. 
        \item[2)] $({\bf e}_r \cdot {\bf e}_\Omega)\neq 0$: \[
                  \cos^2\theta = 1 - \frac{2}{3}\;\left(\frac{r_{\rm co}}{r}\right)^3\] 
     \end{itemize}
    The equilibrium locations described by this solution are chimney-shaped
    surfaces above and below the equatorial plane. This solution only exists for 
    $r \geq (2/3)^{1/3}r_{\rm co}$.

    The left panel of Fig.~\ref{fig_aligned} shows all equilibrium
    locations  for the aligned rotator while the right panel gives
    only the stable equilibria. The `chimney' equilibria are unstable 
    while the equilibria in the equatorial plane are all stable 
    outside the corotation radius.
       
        \begin{figure*} 
        \centerline{\psfig{figure=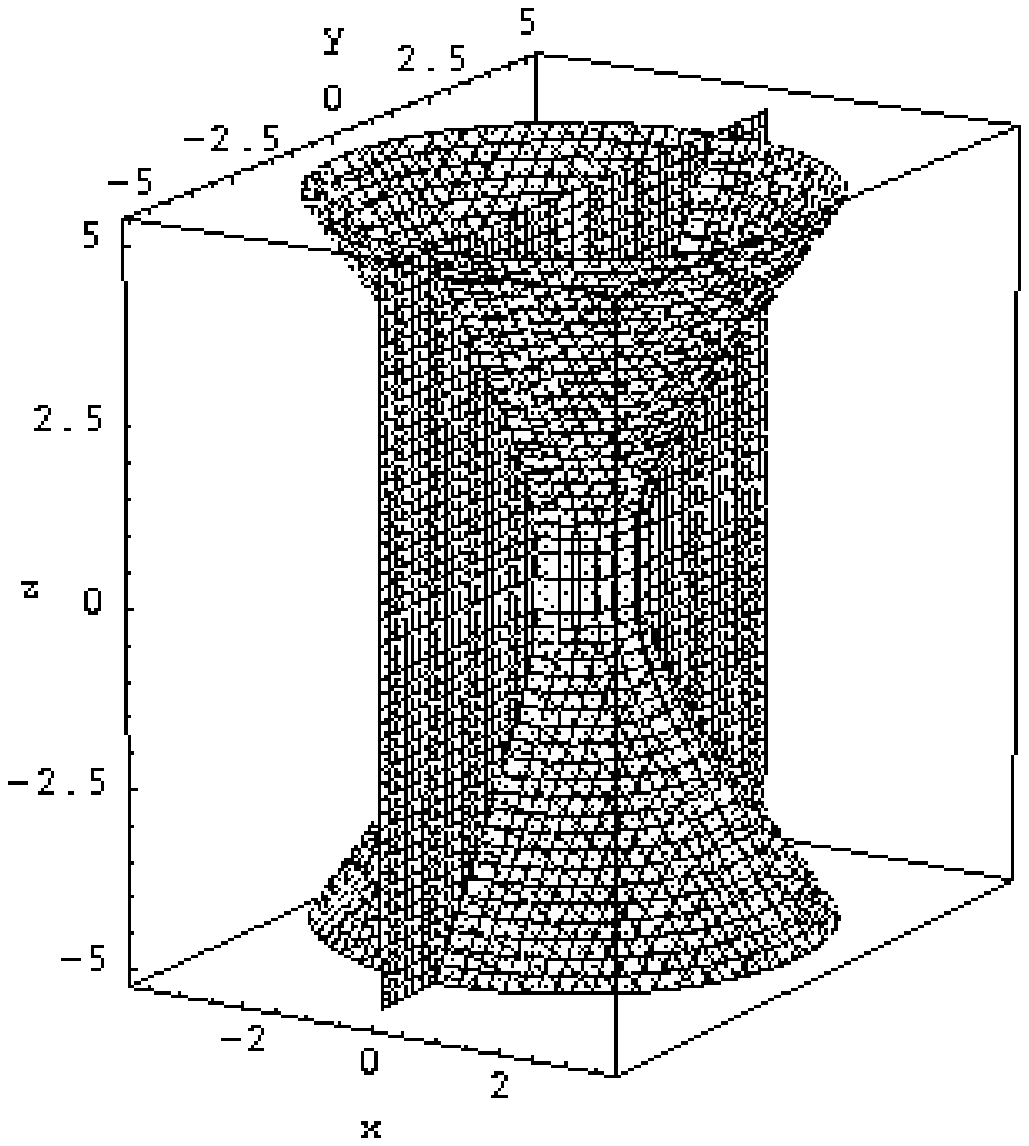,height=2.8in,width=2.9in}
	            \psfig{figure=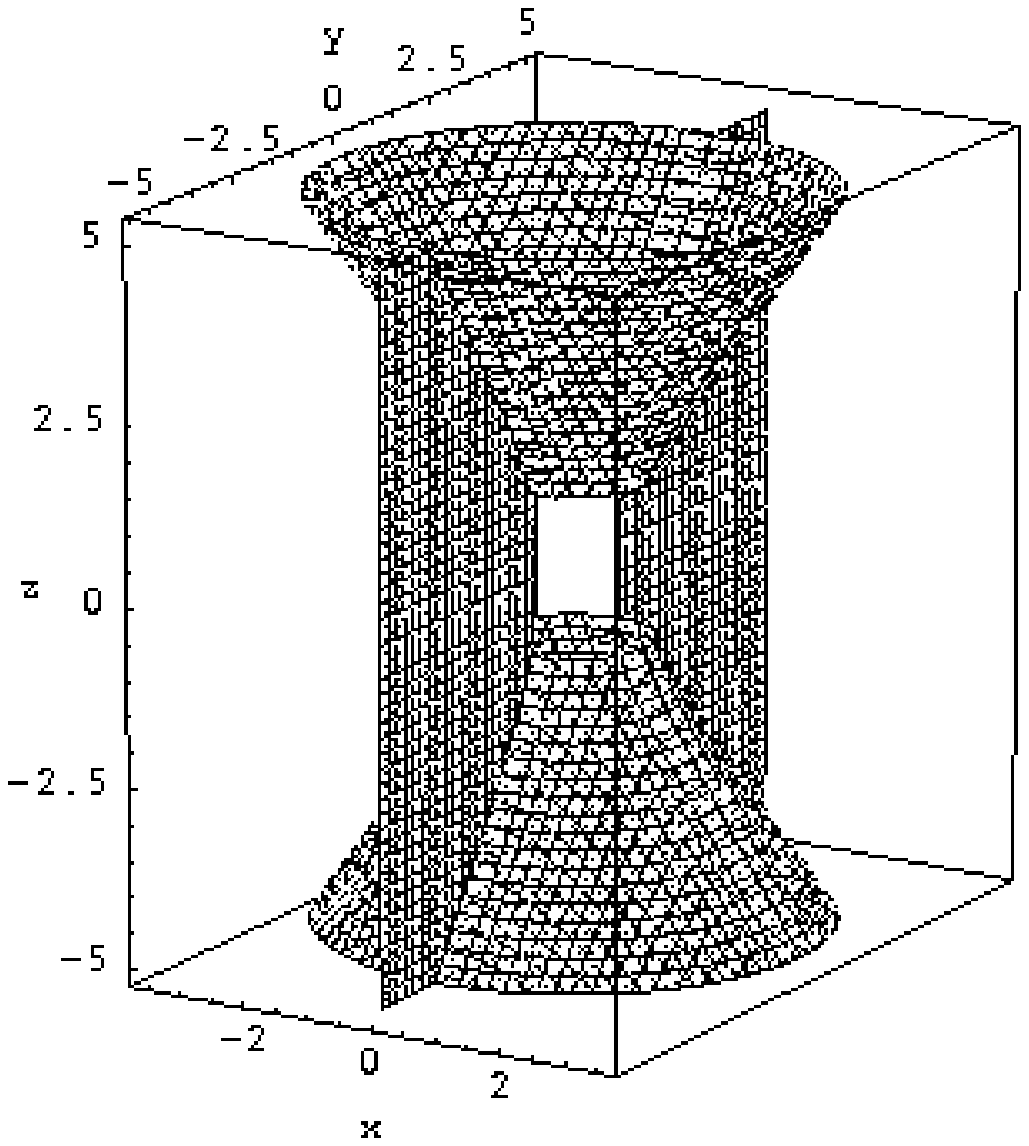,height=2.8in,width=2.9in}}
        \caption{Same as Fig.~\ref{fig_aligned} for the
        perpendicular rotator ($\psi=\pi/2$). The magnetic dipole is oriented 
	along the $x$-axis.}
	\label{fig_perp}  
        \end{figure*}

    \subsection{Perpendicular rotator: $\psi = \pi/2$}  
      
      If the rotation axis and the magnetic axis are perpendicular to
      each other, the equilibrium condition, Eq.~(\ref{equi1}), becomes
      \beq
        \left\{2\left[1-\left(\frac{r_{\rm co}}{r}\right)^3\right]-
	3({\bf e}_r \cdot {\bf e}_\Omega)^2\right\}
        ({\bf e}_r \cdot {\bf e}_{\rm m})=0\;.
      \eeq 
      Similar to the previous case, there are two solutions (see
      Fig.~\ref{fig_perp}): 
      \begin{itemize}
        \item[1)] ${\bf e}_r \cdot {\bf e}_{\rm m}=0$:
                  this solution corresponds to the equatorial plane
                  with respect to the magnetic axis.
        \item[2)] ${\bf e}_r \cdot {\bf e}_{\rm m}\neq 0$:
                  \[\cos^2\theta = \frac{2}{3}\left[1 - \left(\frac{r_{\rm co}}
		  {r}\right)^3\right]\;.\] 
      \end{itemize}
      This corresponds to a chimney structure that is axisymmetric with respect 
      to the rotation axis of the star. The solution only exists for $r \geq r_{\rm co}$.
      
      For $ r \simeq r_{\rm co}$ the co-latitude $\theta$ is about $90^{\circ}$, 
      corresponding to the base of the chimney at the equatorial plane.  For increasing $r$, 
      the co-latitude decreases down to a minimum value of $\theta_{\rm min}= 35.3^{\circ}$
      (for $r \to \infty$) so that at large distances from the star the chimneys form 
      approximately coni with a constant opening angle $\theta_{\rm min}$.

      The stability analysis yields that both equilibrium surfaces are unstable
      in the close vicinity (out to 1-2 $r_{\rm co}$) but are stable further out.

    \subsection{Oblique rotators}

\begin{figure*}[h]
       \centerline{\psfig{figure=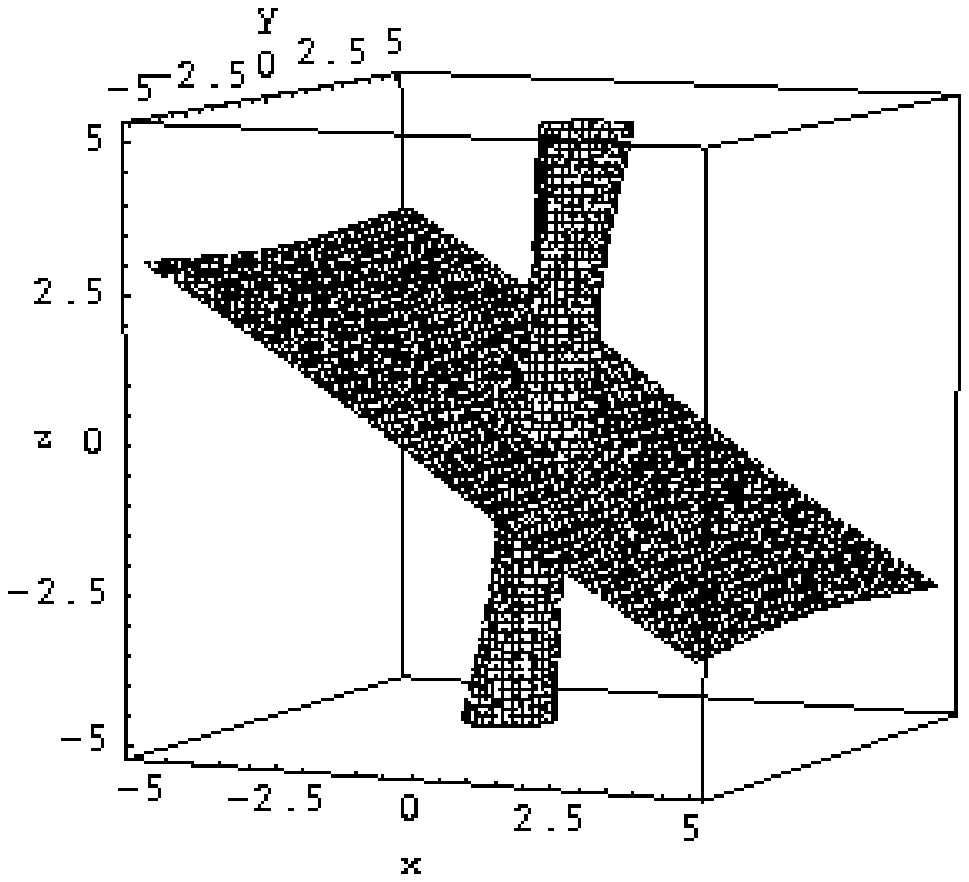,height=2.7in,width=2.7in}
       \hspace{0.3cm}
                   \psfig{figure=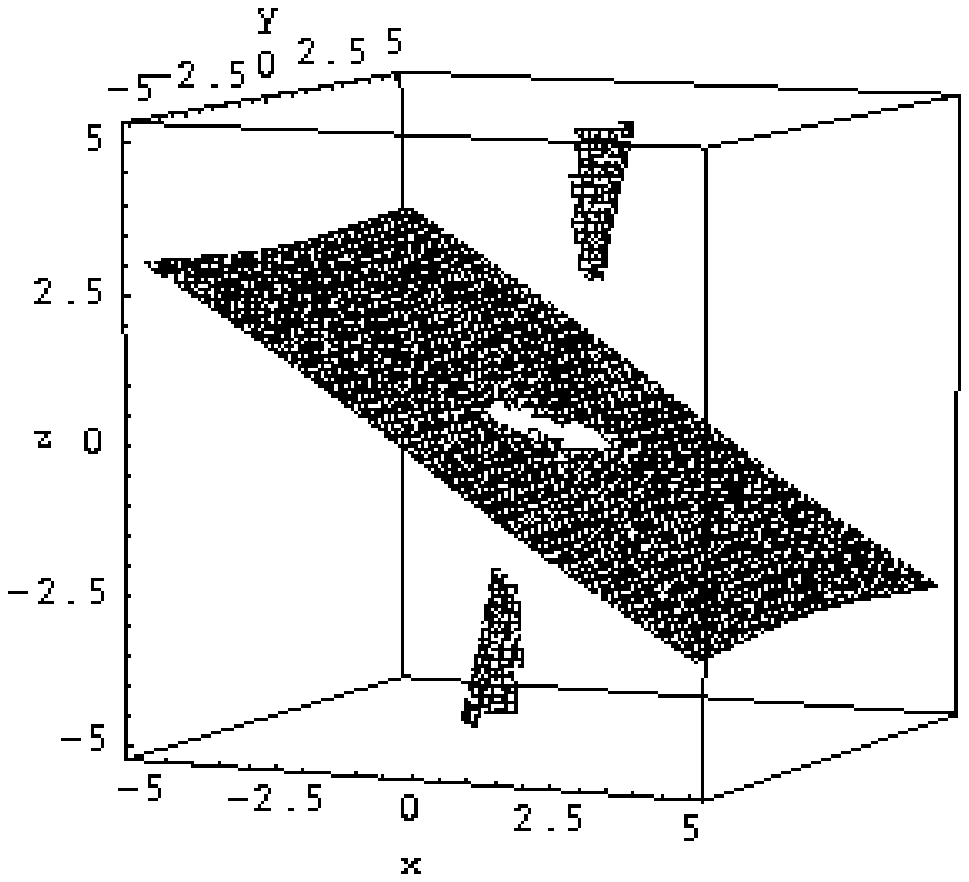,height=2.7in,width=2.7in}}
       \centerline{\psfig{figure=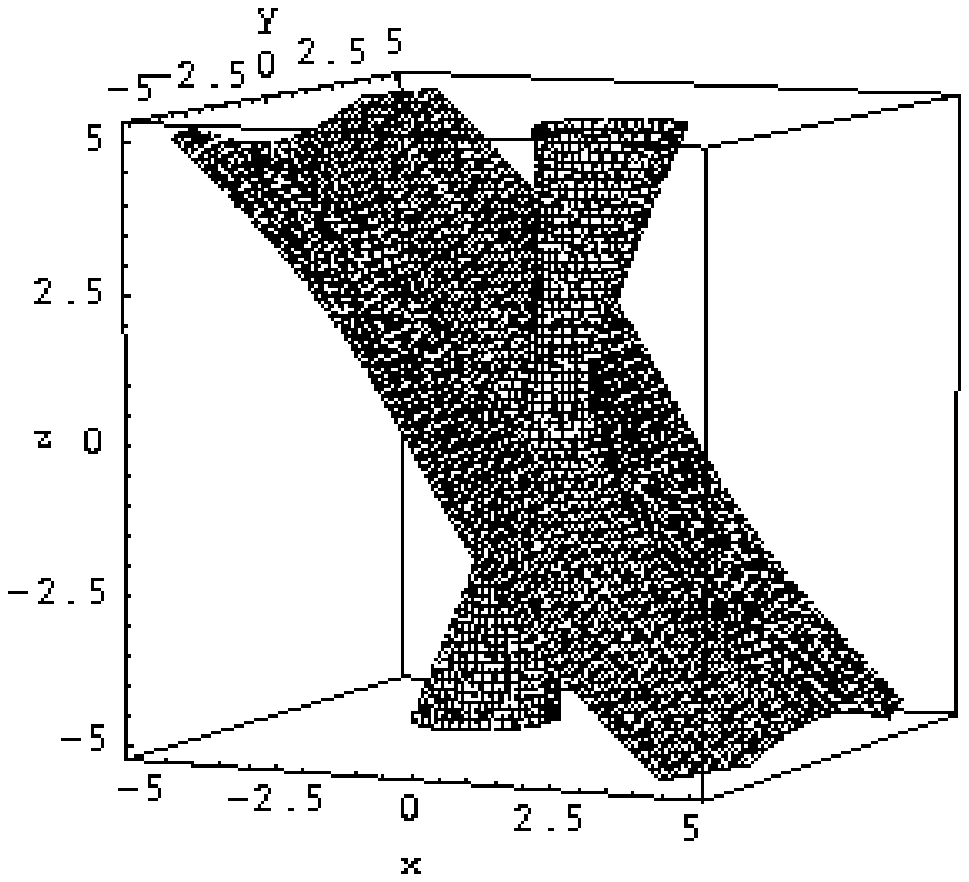,height=2.7in,width=2.7in}
        \hspace{0.3cm}
                   \psfig{figure=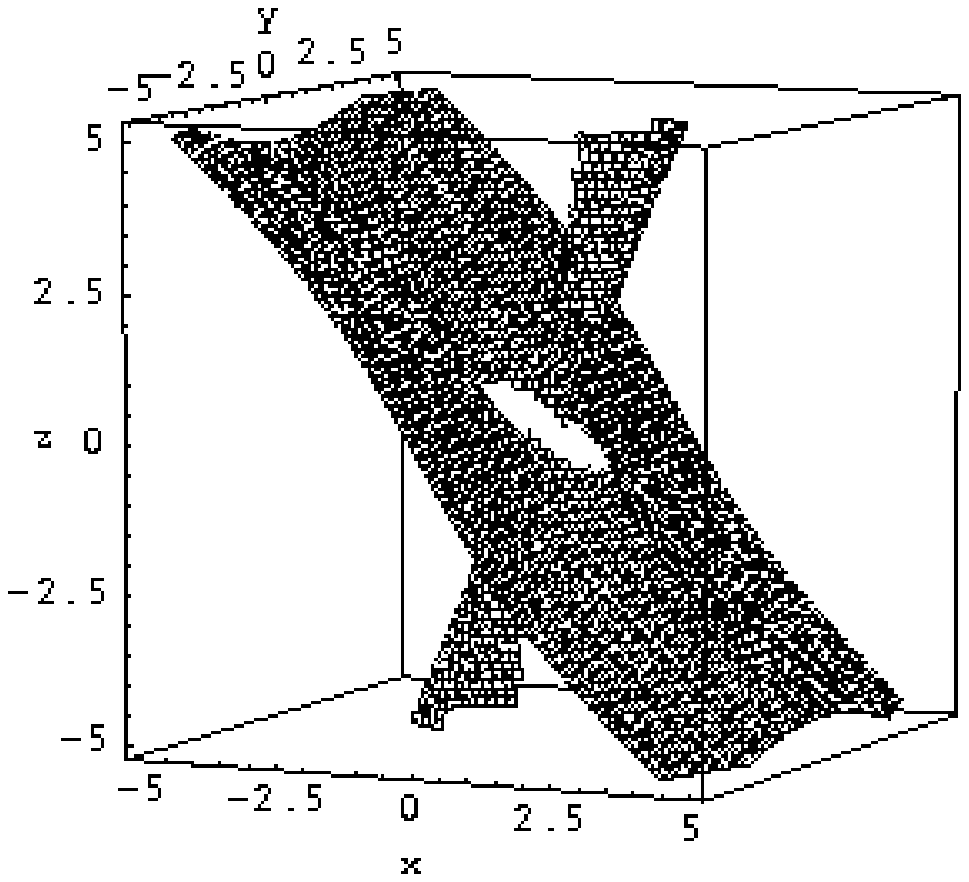,height=2.7in,width=2.7in}}
       \centerline{\psfig{figure=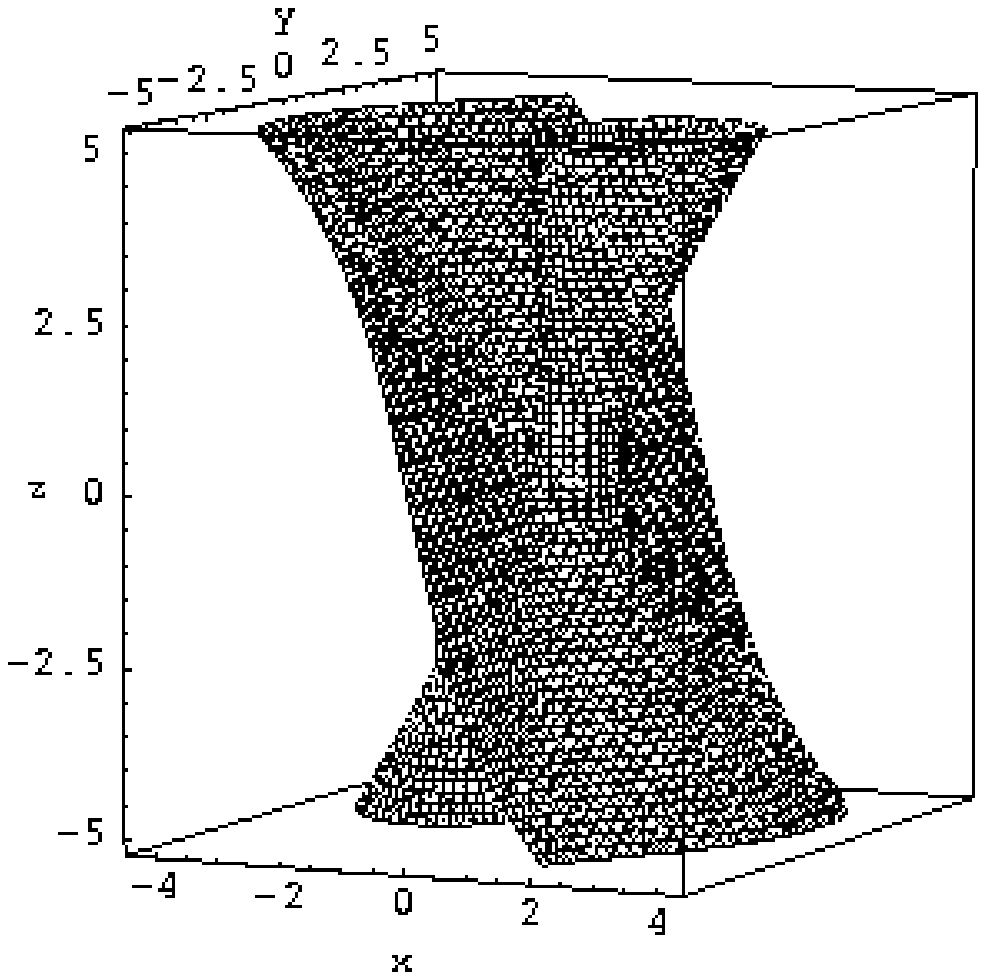,height=2.7in,width=2.7in}
       \hspace{0.3cm}
                   \psfig{figure=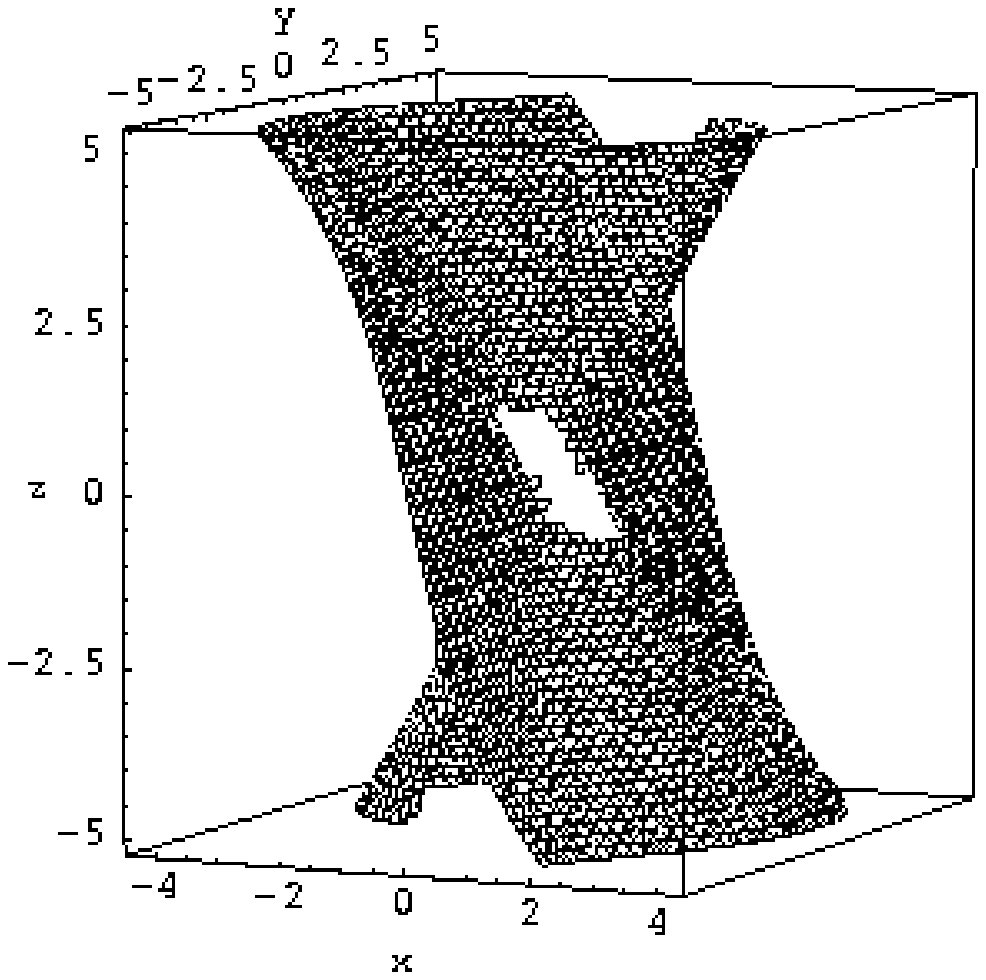,height=2.7in,width=2.7in}}
       \caption{Equilibrium distribution for oblique rotators with various 
       inclination angles (left panels: all equilibrium positions, right panels: only stable 
       positions). Top: $\psi=0.25 \pi$, middle: $\psi=0.375 \pi$, bottom: $\psi=0.45 \pi$.
       The rotation axis is always oriented along the $z$-axis.}
       \label{fig_oblique}
       \end{figure*}

       If the inclination angle between the magnetic dipole and the rotation
       axis is oblique, we have to make use of the complete equilibrium
       condition, Eq.~(\ref{equi1}). Again we can find at least two independent
       solutions which, for small inclinations, describe in good approximation
       a disk-like and a chimney structure. For larger inclination angle,  
       the shapes of the equilibrium surfaces are modified. This can be seen 
       in Fig.~\ref{fig_oblique} for $\psi=0.25 \pi, 0.375 \pi$, and $ 0.45 \pi$, 
       respectively. 
       For intermediate values, $0 \leq \psi\leq 0.5\,\pi$, the disk becomes 
       somewhat warped. In addition to the stable equilibrium points on the warped
       disk, there are also stable regions on the chimneys. Such regions exist 
       for all values of $\psi$, but they move away from the disk for decreasing 
       inclination and reach infinity for $\psi=0$. The size and the curvature of 
       these regions increases with increasing inclination. They are always 
       completely separated from the disk, exept for the case $\psi=\pi/2$ 
       (Fig.~\ref{fig_perp}) where the disk is directly connected to the stable 
       regions on the chimney. In addition we find that the tilt angle $\theta$ 
       between the stable chimney regions and the rotation axis increases continously
       from $\sim 0^{\circ}$ for the aligned rotator up to $35.3^{\circ}$ for 
       the case of the perpendicular rotator. This implies that the new stable 
       equilibrium regions can only exist within a specific angular distance from 
       the rotation axis.              
  
       The approximate alignment of the (warped) disk-like region of equilibria
       with the magnetic equatorial plane can be understood by
       rewriting Eq.~(\ref{equi1}) in the form  
      \beq \label{eq_plane}
       \left\{\left[2\left(1-\left(\frac{r_{\rm co}}{r}\right)^3\right)
       -3({\bf e}_r \cdot {\bf e}_\Omega)^2\right]{\bf e}_{\rm m} +\cos\psi\;
       {\bf e}_\Omega\right\} \cdot{\bf e}_r = 0 \;.
       \eeq

       If the star does not rotate near criticality, we can assume that
       up to some distance from the stellar surface the term in
       angular brackets dominates over the second term. In that limit,
       one solution of Eq.~(\ref{eq_plane}) represents the magnetic
       equator, ${\bf e}_{\rm m}\cdot{\bf e}_r=0$. Consequently, for not
       too rapidly rotating stars, we can always expect an accumulation
       of matter in the magnetic equatorial plane near to the star. However,
       the deviation from a flat disk increases for faster rotating stars
       and is largest in the direction of the tilted dipole. Along
       these longitudes the disk is always bulged towards the rotational equator.

  \subsection{Offset dipoles}

    While we have treated so far only the case of a centered dipole,
    it appears as if this is more an exception than the rule
    for the type of stars we consider (e.g., Neiner 2002). 
    A magnetic dipole with an offset vector ${\bf a}$ relative to the 
    center of the star can be written as 

    \beq\label{boff}
      {\bf B}_{\mbox{\tiny{offset}}} = \frac{\mu_0}{4 \pi \hat{r}^3}\; 
      \left[3({\bf e}_{\bf \hat{r}} \cdot 
      {\bf m}) {\bf e}_{\bf \hat{r}}-{\bf m} \right]\; , 
    \eeq     

    \noindent where ${\bf e}_{\bf \hat{r}}=({\bf r}-{\bf a})/\hat{r}$ and
    $\hat{r}=|{\bf r}-{\bf a}|$.
    The equilibrium positions and their stability are determined in 
    full analogy to the case of centered dipoles.

\begin{figure*}[ht]
      \centerline{\psfig{figure=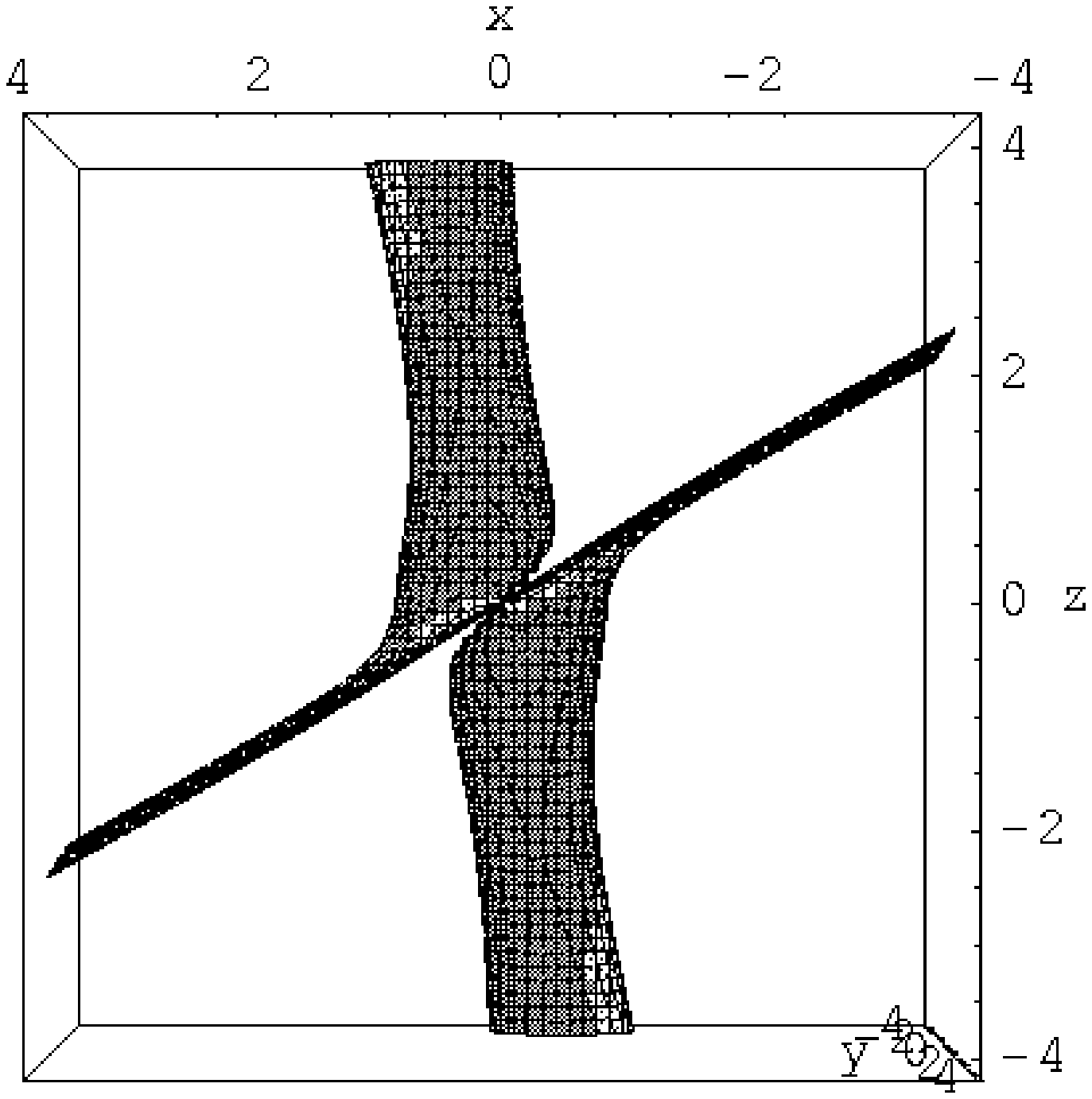,height=2.7in,width=2.7in}
                  \hspace{0.3cm}
                  \psfig{figure=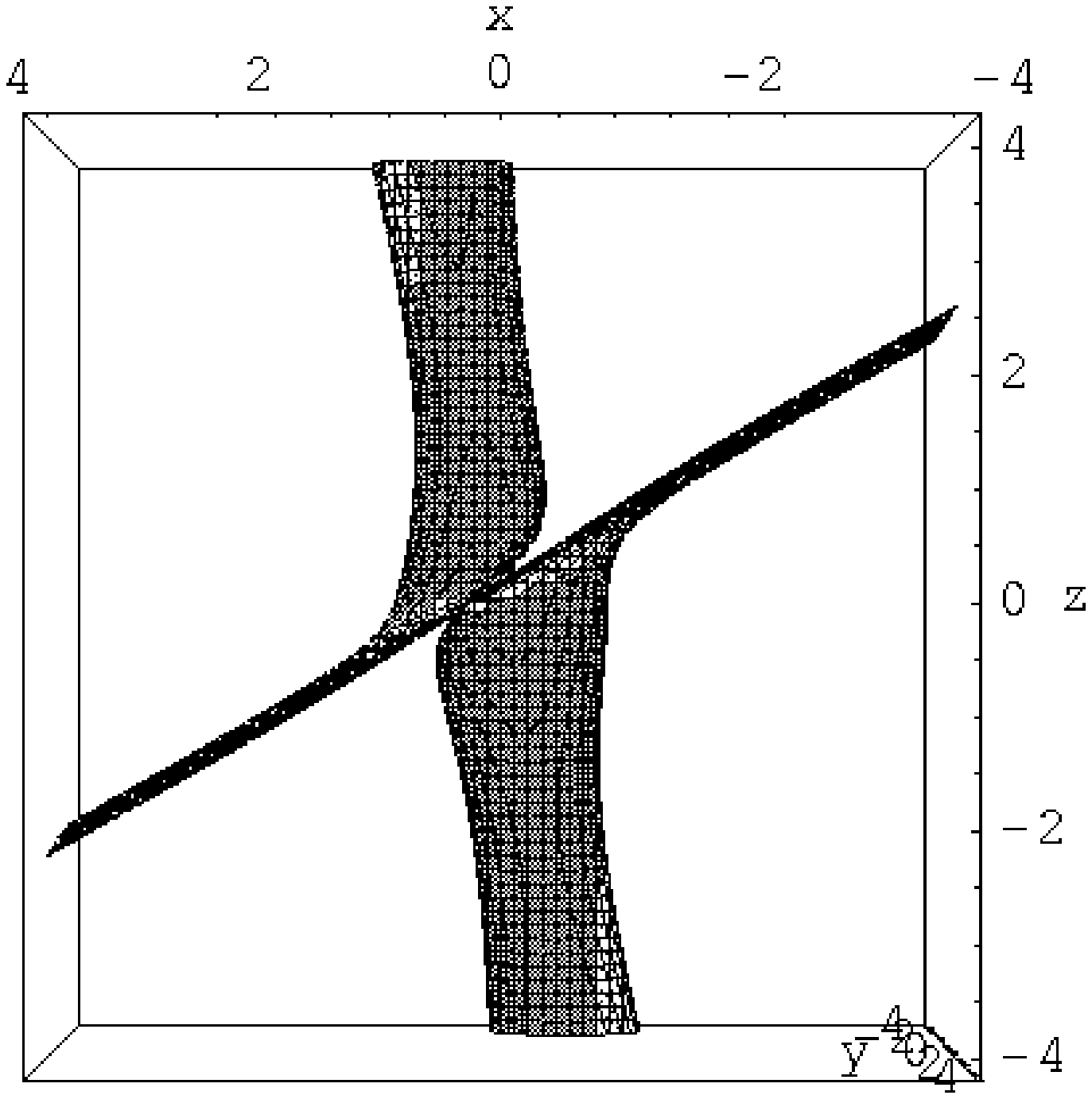,height=2.7in,width=2.7in}}
      \caption{Equilibrium surfaces for inclination angle $\psi=0.25\,\pi$. 
               Left: Centered dipole, right: dipole center displaced by 
	       $0.5R_{\star}$ along the rotation axis ($z$-axis). Areas in 
	       light and dark gray indicate the stable and unstable equilibrium 
	       regions, respectively.}
       \label{fig_offset}
      \end{figure*}

    The qualitative behavior of the equilibrium positions for an offset
    dipole can be seen in Fig.~\ref{fig_offset} for the case $\psi = 0.25\,
    \pi$ in comparison with the corresponding case with a centered dipole. 
    The stellar parameters are the same as before.    
    In the right panel of Fig.~\ref{fig_offset} the dipole has been moved by
    $0.5\,R_{\star}$ along the rotation axis ($z$-axis). The shape 
    of the equilibrium distribution is only slightly modified in comparison to 
    the centered dipole. This is also the case for other realistic offsets
    (i.e., such that the dipole center remains within the star). Taking into 
    account that we are mainly interested in the large scale distribution of
    matter $(\sim 5-6\,\, r_{\rm co})$ this result is not surprising since the 
    field geometry on this scale is hardly influenced by dipole shifts 
    $\ll 1\,r_{\rm co}$. Consequently, changes in the equilibrium distribution 
    of circumstellar matter due to reasonable magnetic offsets are almost 
    negligible.

  \section{Discussion}
  \label{sec_disc}
    The origin of corotating circumstellar matter around hot
    stars with strong magnetic fields is commonly explained
    with the models of Cassinelli et al. (2002) and
    Babel \& Montmerle (1997) in terms of a radiatively driven 
    stellar wind following the (approximately dipolar) magnetic field 
    lines. Since the material streams from both hemispheres of 
    the stellar surface and reaches supersonic speed, the model 
    suggests that the plasma collides and forms a shock in the 
    magnetic equator. The plasma cools through UV and X-ray emission  
    and accumulates in an equatorial disk. While most of the theoretical 
    calculations assume the special case of aligned magnetic and
    rotation axes, we have determined the equilibrium plasma distributions 
    for the general case of an oblique dipolar rotator and also 
    determined their stability properties. For aligned magnetic and 
    rotation axes we confirm the accumulation of matter in a circumstellar 
    disk. For increasing inclination angle the disk becomes warped and 
    additional stable equilibrium positions appear. These are curved 
    surfaces, roughly aligned with the axis of rotation, 
    located above and below the magnetic equatorial plane. They grow 
    and move towards the star for increasing values of $\psi$.     
    For the case of a decentered dipole we find that for reasonable 
    offsets the stable equilibrium positions are 
    changed only slightly compared with a centered dipole. We expect that,
    analogous to the formation of the equatorial disk, the 
    newfound stable equilibrium positions lead to the local 
    accumulation of plasma. Incoming material from the wind is cooled 
    down by radiation behind shocks. Therefore, depending on the kinetic 
    energy of the incoming plasma, we expect that the mass accumulation
    outside the disk appear in observations of continous and rotationally 
    modulated IR, UV and X-ray emission, which is affected by the size 
    of the emission region and its relative orientation to the line of sight.

    In fact, there are observations which indicate the existence of additional
    equilibrium positions outside the equatorial disk. For the rapidly 
    rotating Be star $\omega$ Orionis, Neiner et al. (2003) found evidence 
    for the presence of two corotating regions of material outside the plane
    of the disk by measuring variations in the peak intensities of emission 
    lines. Such clouds were also reported by Short \& Bolton (1994) for the 
    magnetic He-strong star $\sigma$ Ori E. 
    
    We are aware that our model is rather crude. Possible extensions include
    the dynamics of the radiatively driven wind and the corresponding cooling
    and accumulation processes. These should be incorporated into three-dimensional 
    MHD simulations in order to understand the dynamics and distribution of 
    plasma in rotating dipolar fields. First steps in this direction are the 
    two-dimensional MHD simulations by Ud-Doula \& Owocki (2002) (see also Ud-Doula 
    \& Owocki, 2003).
      
  \section{Conclusion}
  \label{sec_conc}
    A simple consideration of the mechanical equilibrium of plasma in the
    magnetosphere of a rotating magnetic star with relative inclination
    of the magnetic and rotation axes suggests the following distribution
    of matter accumulating, e.g., from a stellar wind:
    \begin{enumerate}
      \item a disk-like structure roughly aligned with the magnetic equatorial
            plane
      \item two locations above and below the disk, coarsely aligned
            with the axis of rotation.	    
    \end{enumerate}
    The latter stable equilibrium regions are most prominent for large obliquity
    of the axes (perpendicular rotator). They provide hitherto unknown 
    locations for the accumulation of circumstellar matter, which could
    explain some observations of rapidly rotating magnetic stars, e.g. the 
    close alignment of the new stable regions with the rotation axis could result
    in an observable rotational modulation of the disk radiation.

     {\em  \noindent We are grateful to Matthias D. Rempel and Francisco Frutos
           Alfaro for valuable discussions and to Coralie Neiner for pointing us to
           $\omega$ Orionis and $\sigma$Ori E.\\ This research has made use
           of NASA's Astrophysics Data System Abstract Service.}

\clearpage

\end{document}